\documentclass[aps,prb,floatfix,twocolumn,superscriptaddress,showpacs,amsmath,amssymb]{revtex4}
\usepackage{graphicx}
\usepackage{multirow}
\usepackage{xcolor}
\usepackage{fancyhdr}

\newcommand{\udarrow}{{\uparrow/\downarrow}}
\newcommand{\lapl}{\nabla^2}

\makeatletter
\def\blfootnote{\gdef\@thefnmark{}\@footnotetext}
\makeatother

\begin{document}

\title{
An efficient hybrid orbital representation for quantum Monte Carlo calculations
\blfootnote{
This manuscript has been authored by UT-Battelle, LLC under Contract No. DE-AC05-00OR22725 with the U.S. Department of Energy.  The United States Government retains and the publisher, by accepting the article for publication, acknowledges that the United States Government retains a non-exclusive, paid-up, irrevocable, world-wide license to publish or reproduce the published form of this manuscript, or allow others to do so, for United States Government purposes.  The Department of Energy will provide public access to these results of federally sponsored research in accordance with the DOE Public Access Plan(http://energy.gov/downloads/doe-public-access-plan).
}
}

\author{Ye Luo}
\email{yeluo@anl.gov}
\affiliation{Argonne Leadership Computing Facility, Argonne National Laboratory, Argonne, Illinois 60439 U.S.A.}

\author{Kenneth P. Esler}
\affiliation{Stone Ridge Technology, 2015 Emmorton Rd. Suite 204, Bel Air, Maryland 21015 U.S.A.}

\author{Paul R. C. Kent}
\affiliation{Center for Nanophase Materials Sciences and Computational Sciences and Engineering Division, Oak Ridge National Laboratory, Oak Ridge, Tennessee 37831 U.S.A.}

\author{Luke Shulenburger}
\affiliation{HEDP Theory Department, Sandia National Laboratories, Albuquerque, New Mexico 87185 U.S.A.}
\date{\today}


\begin{abstract}
The scale and complexity of quantum system to which real-space quantum Monte Carlo (QMC) can be applied in part depends on the representation and memory usage of the trial wavefunction. B-splines, the computationally most efficient basis set, can have memory requirements exceeding the capacity of a single computational node. This situation has traditionally forced a difficult choice of either using slow internode communication or a potentially less accurate but smaller basis set such as Gaussians.  Here, we introduce a hybrid representation of the single particle orbitals that combine a localized atomic basis set around atomic cores and B-splines in the interstitial regions to reduce the memory usage while retaining high speed of evaluation and either retaining or increasing overall accuracy.  We present a benchmark calculation for NiO demonstrating a superior accuracy while using only one eighth the memory required for conventional B-splines. The hybrid orbital representation therefore expands the overall range of systems that can be practically studied with QMC.
\end{abstract}

\maketitle

\section{Introduction}

Continuum quantum Monte Carlo (QMC) methods~\cite{rev:qmcsolids,Needs2010} are among the most accurate for solving the many-body Shr\"{o}dinger equation with an ab initio Hamiltonian. They have been applied to a wide range of materials from molecules~\cite{Benali2014,Krongchon2017,Zen2018} and clusters~\cite{Zen2016,Mouhat2017,Motta2017,Fumanal2016} to solids~\cite{Chen2016,Luo2016,Tsatsoulis2017,Azadi2017,Trail2017,Varsano2017,Mazzola2018,Shin2017,Kylanpaa2017,Dzubak2017,Santana2017,Yu2017,Saritas2017}, to non-covalently bonded systems~\cite{Dubecky2016} and to systems with strong electronic correlation~\cite{Wagner2016}. Among them, variational~\cite{McMillan1965} and diffusion Monte Carlo~\cite{Grimm1971,Kalos1974} (VMC, DMC) have been used to study large systems with more than a thousand electrons~\cite{Hood2012,Luo2016}.

In these calculations, importance sampling using a trial wavefunction is employed for both accuracy and efficiency. The evaluation of single particle orbitals (SPO) in the trial wavefunction is an important component of the overall computational cost. The conventional representations for SPOs using either plane waves (PW), a linear combination of atomic orbitals (LCAO), or a real space representation such as B-splines~\cite{Alfe2004,parker2015} for the underlying basis set. PWs can be readily transformed to B-splines. The basis set must be large enough to accurately represent the orbitals with sufficient accuracy and is ideally fast to evaluate on modern computer systems with a minimum of numerical operations. Basis sets that are readily converged via a single parameter such as plane-wave cutoff energy or grid spacing are preferred due to their facile ease of use. Today, B-spline approaches are preferred for solids, some molecular calculations, and are an ideal choice for surfaces. They require a fixed number of elements to be evaluated for each orbital independent of system size. However, for sufficient accuracy they must be stored on a sufficiently fine grid, which requires a significant amount of memory in practice.  This large amount of memory cannot be easily distributed among several nodes due to the frequent and random access pattern and it quickly exhausts the on-node memory in challenging simulations even when fully shared by threads and processes within the node. Use of symmetry can reduce memory usage of all basis sets, but this avenue is not available for many scientifically rich problems such as those involving defects and dopants or surfaces of realistic materials. The use of heavier elements also tends to require a finer grid, dramatically increasing memory consumption. For the above reasons, approaches enabling a smaller memory footprint without compromising the accuracy or speed of evaluation are needed. Note that these are distinct considerations from those required by quantum chemical and density functional calculations where the ability to perform numerical integrals over basis functions is paramount. Here, rapid and accurate evaluation of the basis functions at specific points in space is paramount.

At least three approaches to reduce memory usage have been devised: (i) Using real-space localization of the orbitals and truncation to reduce the grid sizes\cite{Reboredo2005,Alfe2004-lo}. This saves a system-dependent but significant amount of memory, particularly for systems with vacuum regions, but requires the truncated orbitals to be constructed; (ii) Using a mixed or hybrid approach~\cite{Esler2012} with a combined atomic basis set and B-splines. Within atom-centered spheres, the atomic basis reproduces the most rapidly varying part of the orbitals, while the B-splines reproduce the more slowly varying parts. Continuity at the sphere boundaries must be preserved which either requires special care during orbital generation or when choosing the B-spline grid. The former is severely limited by the availability of such methods and the later limits the memory savings; (iii) Classifying the orbitals into different groups, e.g. via their kinetic operator~\cite{Krogel2018}, and using different tailored B-spline grids for each group. The savings, demonstrated up to 60\%, are system dependent and smaller than the previous two schemes, but could be used in conjunction with them.

In this work, we improve on the second approach and develop a hybrid orbital representation which combines treatment of atomic regions using spherical harmonic expansions while retaining coarse 3D B-splines to efficiently represent the interstitial region. Our approach does not impose any requirements on input orbitals, and could be combined with the other approaches for further efficiency increases. Unlike the earlier development of a similar basis set~\cite{Esler2012} we explicitly treat the continuity at the boundaries of regions for flexibly coarsening the B-spline grid to maximize memory savings. In an example solid-state NiO benchmark, we show the necessary steps to control the sizes of the spherical basis sets and radii of atomic regions, and the high accuracy, evaluation speed and memory savings that can be achieved.

\section{Computational methods}
\label{sec:method}
\subsection{Trial wavefunctions in QMC}
A high-quality trial wavefunction plays a key role in QMC simulations. In VMC, it directly determines the result, while in DMC, it not only enables importance sampling and governs the fixed node error, but also controls the pseudopotential locality error~\cite{Dzubak2017a,Krogel2017} when non-local pseudopotentials~\cite{Drummond2016,Krogel2016,Trail2017pp} are used.
The most used many body trial wavefunction ansatz is the Slater-Jastrow form
\begin{align}
  \Psi(\mathbf{r}) &= \left(\sum_m D^\uparrow_m D^\downarrow_m\right) \exp(J(\{\mathbf{r}\})) \\
  D^\udarrow & = \left|
    \begin{array}{ccc}
      \phi_1(\mathbf{r}_1) & \cdots & \phi_n(\mathbf{r}_1) \\
      \vdots    & \ddots & \vdots \\
      \phi_1(\mathbf{r}_n) & \cdots & \phi_n(\mathbf{r}_n) \\
    \end{array}
  \right|,
\end{align}
where $D^\udarrow_m$ are Slater determinants built with SPOs $\phi_i$ computed at given electron coordinates $\{\mathbf{r}_i\}$, and the Jastrow factor $J$ explicitly builds in electron correlations.

\begin{figure}[t]
\centering
\includegraphics[width=\columnwidth]{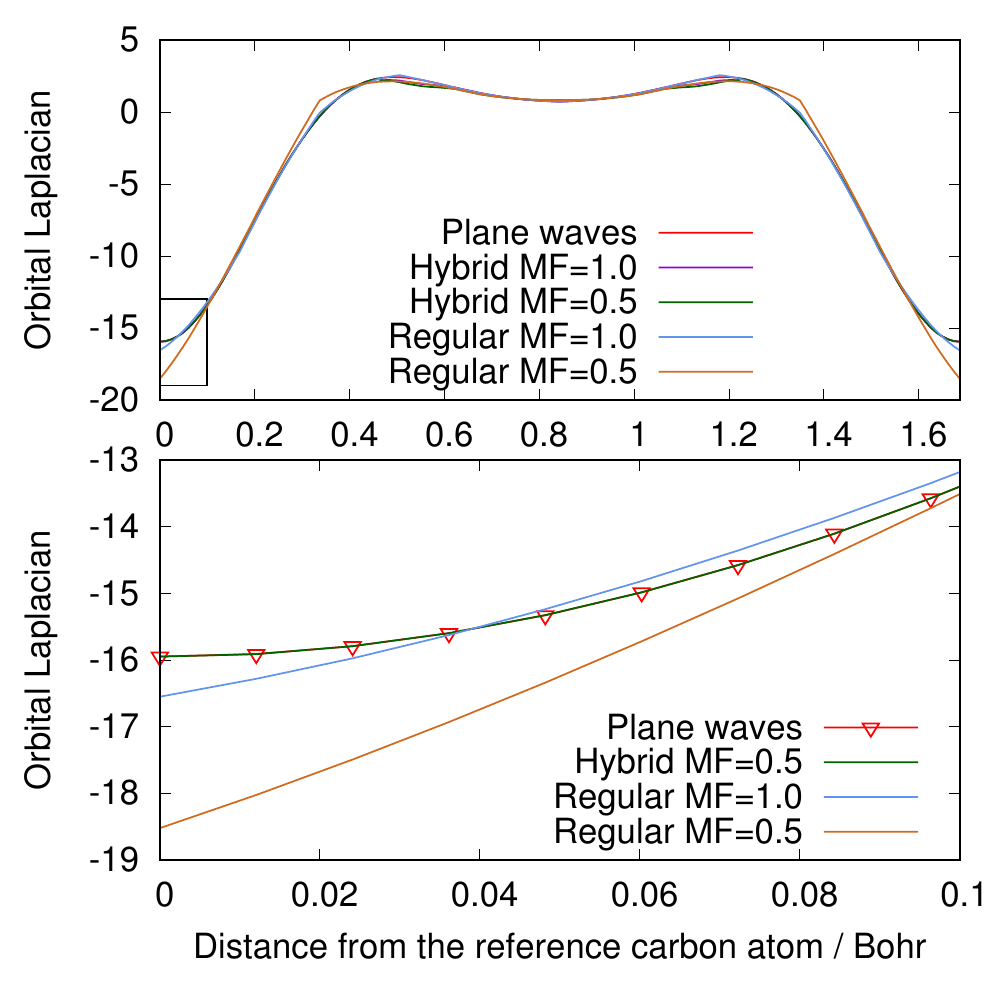}
\caption{Laplacian of the 2S bonding orbital in carbon diamond scanned between the two carbon atoms within a primitive cell.
 The largest error in the regular representation occurs near the cores.  The hybrid orbital representation eliminates that error and gives the same accuracy
 as the original orbital described by plane waves from DFT.  Mesh-factor (MF) describes the coarseness of the spline grid, with a MF = 0.5 being the same number of points as the PW representation and MF = 1.0 having twice as many points in each direction.  Memory usage scales as MF$^3$.}
\label{fig:orbitals}
\end{figure}

SPOs are usually generated by mean field methods such as density functional theory~\cite{Hohenberg1964,Kohn1965} (DFT) or Hartree-Fock and they can be expressed in LCAOs or PWs. LCAO is extremely compact but its completeness is usually inferior to PW. LCAO requires choosing a proper atomic basis set while PW only requires controlling a single kinetic energy cutoff. However, neither representation is ideal for QMC which requires repeatedly evaluating the orbital value at arbitrary points in real space.  Evaluating both LCAO and PW representations cost $N_{\rm basis}N_{\rm orbs}N_{\rm elec}$ where all three of these grow with the number of electrons unless truncation is applied for the atomic orbitals.  As an alternative to these, QMCPACK\cite{qmcpack2018} expresses SPOs with tricubic B-splines (also referred as `regular' in the rest of this work) which have a much lower computational cost $64 N_{\rm orbs}N_{\rm elec}$ ($N^{\rm LCAO}_{\rm basis}>64$, $N^{\rm PW}_{\rm basis}\gg 64$) due to their local support i.e. fixed evaluation cost per electron independent of system size or atom count. In real-space QMC codes, the spline coefficients are usually replicated a single time on each node for fast random access. This limits the possibility for large simulations unless the nodes have a very large amount of memory.
Fig.~\ref{fig:orbitals} demonstrates that the accuracy is reduced when a coarse grid is used in B-spline representation.
A compromise between memory and accuracy often has to be made for large system sizes or complex structures.

\subsection{Hybrid orbital representation}
To minimize the memory usage vs accuracy dilemma of B-splines, we introduce a hybrid orbital representation to achieve high accuracy, improve compactness and retain high evaluation speed. Since orbitals near nuclei are often atomic-like, they can be compactly represented via radial expansions in spherical harmonics.  For regions far away from nuclei, slowly varying representations such as B-splines are a better choice. Our hybrid orbital representation combines these two representations. These are similar considerations to those motivating the historical development of augmented plane wave (APW) and linearly augmented plane wave (LAPW) approaches in electronic structure.\cite{MartinBook2004}

The idea of a hybrid orbital representation in QMC~\cite{Esler2012} was originally introduced to overcome memory limitations imposed by graphical processing units (GPUs). However, the continuity at the boundary was not explicitly treated. Thus the SPOs either had to be generated by the LAPW method where the orbital values at the boundaries between atomic and interstitial regions are matched by construction, or by plane waves, but using a rather conservative choice of grid coarseness to retain continuity at the boundary.
Our new approach requires only the usual PW DFT calculation and also allows adjusting the grid size freely.

\begin{figure}[t]
\centering
\includegraphics[width=0.5\columnwidth]{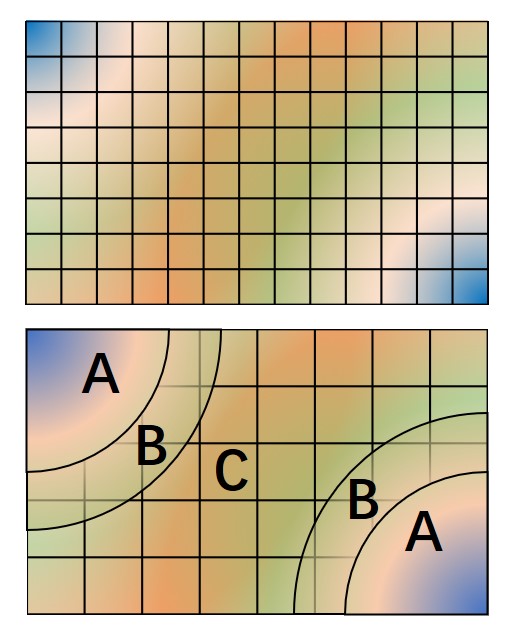}
\caption{Illustration of regular and hybrid orbital representation. Regular B-spline representation (upper panel) contains only one region and a sufficiently fine mesh to resolve orbitals near the nucleus. The hybrid orbital representation (lower panel) contains near nucleus (A) regions where spherical harmonics and radial functions are used, buffers (B) or interpolation regions, and an interstitial (C) region where a coarse B-spline mesh is utilized.}
\label{fig:threeregions}
\end{figure}

In our new approach, the whole space is divided into near nucleus (A), buffer or interpolation (B), and interstitial (C) regions, as shown in fig.~\ref{fig:threeregions}. In region A, orbitals are computed as
\[
  \phi^A_n({\bf r})=R_{n,l,m}(r)Y_{l,m}(\hat{r})
\]
Where $Y_{l,m}$ are the real-valued spherical harmonics and $R_{n,l,m}$ is the radial function for the $n$-th orbital on the $Y_{l,m}$ basis. Even for a complex orbital, $\phi_n$, $R_{n,l,m}$ can be treated as a double sized real array to maximize the computational efficiency by enabling vectorization on the CPU. Since radial functions are not provided as an input, they are generated by transforming the PW coefficients of a given set of SPOs with the spherical harmonic addition theorem. The radial functions are expressed by 1D B-splines on a uniform mesh with a default 0.02\,bohr spacing. This step can be extremely computationally intensive and we provide a very efficient way in appendix~\ref{app:parallel}.  QMC algorithms also require computing the gradients and Laplacian of $\phi$ for measuring kinetic energy, and the formulas are provided as appendix~\ref{app:gradlapl}. In region C, the orbitals are evaluated as regular tricubic B-splines.

The B region is designed to enforce the continuity of value, first and second derivatives of the orbitals between the A and C regions and alleviate the numerical pathology that can arise due to discontinuities between the regions.  These discontinuities can arise for several reasons.  First, orbitals are expanded in two distinct incomplete basis sets and are unlikely to match exactly except when the basis sets are very large. Second, the B-splines exactly match the PW reference only on the mesh grid points and interpolation between grid points causes mismatching, particularly as higher frequency components become more important near atoms.  Finally, intentionally coarsening the mesh to save memory exacerbates this discrepancy on a direct A/C boundary. Therefore, we introduce a region B, where orbitals are computed as
\begin{align}
  \phi^B_n({\bf r}) &= S(r) \phi^A_n({\bf r}) + (1-S(r))\phi^C_n({\bf r}) \\
                S(r) &= \frac{1}{2}-\frac{1}{2}\tanh\left[\alpha\left(\frac{r-r_{\rm A/B}}{r_{\rm B/C}-r_{\rm A/B}}-\frac{1}{2}\right)\right]
\end{align}
where $r_{\rm A/B}$ and $r_{\rm B/C}$ are the distances from boundary to the nucleus. $S(r)$ is a smooth function with a parameter $\alpha$ controlling the strength of the smoothing and continuity of first and second derivatives of $\phi_n({\bf r})$. The function form is chosen to have values varying from approximately one at $r_{\rm A/B}$ to approximately zero at $r_{\rm B/C}$ while the first and second derivatives are approximately zero at both ends. The larger $\alpha$, the smaller discrepancy on the boundary and more rapid change inside buffer region. $\alpha=2$ and $r_{\rm B/C}-r_{\rm A/B}=0.3\,{\rm bohr}$ are found to be suitable for the Ni and O atoms in our NiO test case as well as the C atom in the carbon diamond and are thus kept fixed for the whole study.

In the following sections, we will study the choices of the largest angular momentum ($l_{\rm max}$), radial function cutoff ($r_{\rm B/C}$) for each species and mesh-factor (MF) of interstitial region in NiO via variational Monte Carlo. Then, we demonstrate the advantages of using hybrid orbital representation in diffusion Monte Carlo. All data needed to reproduce the calculations and analyses is available at \url{https://materials_data_facility_url_available_after_acceptance}.

\section{Results}
\label{sec:results}
\subsection{NiO supercell}
We choose a 32 atom antiferromagnetic supercell of crystalline B1 NiO as our benchmark system. In our calculations, the core electrons, (1s,2s,2p) of Ni and (1s) of O, respectively, are removed by using scalar relativistic norm-conserving pseudopotentials(PPs) generated by DFT-LDA atomic calculations and customized for QMC by using very small cutoff radii for a more accurate core description.\cite{Krogel2016,Root2015} The SPOs are generated initially with Quantum ESPRESSO~\cite{qe:main,Giannozzi2017}.  The plane-wave cutoff in DFT calculations was 200 Ha. This is much higher than conventional DFT calculations due to the use of a very hard Ni pseudopotential with semi-core electrons in the valence. The self-consistent total energies were converged to 0.01~mHa/f.u.\ with respect to this cutoff.  The Jastrow part of the trial wave function contains both one- and two-body Jastrow factors with a total of 40 parameters optimized by energy minimization~\cite{Umrigar:linear} within VMC.  The one-body Jastrow has a 4.8262\,bohr cutoff radius while the two-body Jastrow cutoff is 5.5728\,bohr the same as the Wigner-Seitz radius of the supercell. These parameters are similar to those of a recent NiO QMC study.\cite{Shin2017} All our QMC calculations are performed with a modified version of QMCPACK~\cite{qmcpack2018}. The NiO supercell is a test case included in QMCPACK release.

\subsection{Choosing largest angular momentum, cutoff radius and mesh-factor with VMC}
When using the hybrid orbital representation, the quality of the wavefunction is controlled by three factors: (1) the number of angular momentum channels used in region A, (2) the cutoff radius for region A and (3) the mesh-factor in region C. In QMCPACK, the mesh-factor (MF) is set to one by default, yielding a grid as large as the charge density grid in the PW DFT calculation, determined by the PW cutoff. With a MF not equal to one, the number of grid point in each direction is scaled by MF. There is a trade-off between using large cutoff radii which will require fewer B-spline points (smaller mesh-factors) but will need more angular momentum channels and thus higher computational cost to remain accurate.  Conversely, small cutoff radii may need fewer angular momentum channels, but will require finer B-spline grids. Additionally, the size of region B should be sufficiently large to have a good smoothing but it can not be too large due to the potential inaccuracy of either representation. Using a 0.3\,bohr thick region inside the cutoff radius was appropriate in our test cases. We do not anticipate significant sensitivity to this choice provided the region is kept similar in scale to the B-spline mesh.

\begin{figure}[t]
\centering
\includegraphics[width=\columnwidth]{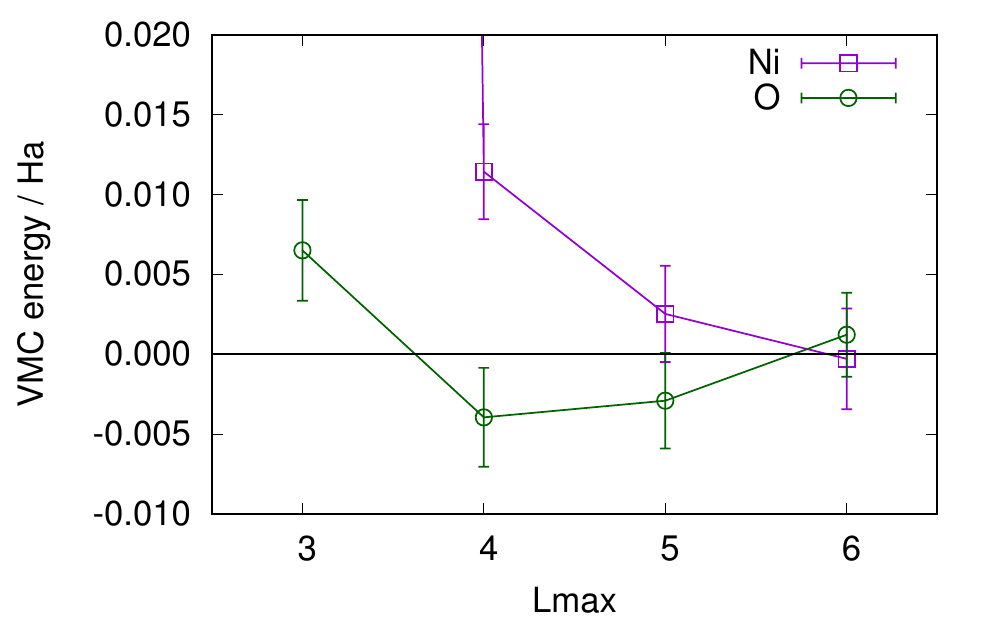}
\caption{The VMC energy convergence of the largest angular momentum used for each species. $L_{\rm max}=7$ is used as a reference. $L_{\rm max}=5,4$ for Ni and O atoms respectively give errors of 3(3) and 4(3) mHa in the test case of 32 atom NiO supercell.}
\label{fig:lmax}
\end{figure}

We first study the accuracy impact of the largest angular momentum $l_{\rm max}$ for each species. When $l_{\rm max}$ is larger, the basis available to describe the orbitals is more complete but this increases computational cost, which grows as $(l_{\rm max}+1)^2$, as well as memory storage. In principle, $l_{\rm max}$ is entangled with the cutoff radius $r_{\rm B/C}$.  At large distances from the ion, orbitals are much less atomic-like and need a larger basis to recover an accurate description.  In practice, it is possible to achieve convergence of the basis over a wide range of $r_{\rm B/C}$.  Here we study the affect of $l_{\rm max}$ on a per species basis with $r^{\rm Ni,O}_{\rm A/B}=r^{\rm Ni,O}_{\rm B/C}=1.0$ by computing the VMC energy using different choices for the maximum angular momentum near each atom.  When we vary $l^{\rm max}$ for one species, it is set at 7 for the other to ensure any imperfection in the wavefunction only comes from the species being studied.  Since Ni has 3-$d$ semi-core electrons as valence, $l_{\rm max}$ is scanned from 3 to 7 shown in Fig.~\ref{fig:lmax} and we find $l^{\rm Ni}_{\rm max}=5$ and $l^{\rm O}_{\rm max}=4$ gives total energies consistent with a fully converged reference calculation ($l_{\rm max}=7$) within 2 standard deviations. These settings are used for the rest of this study.

\begin{figure}[t]
\centering
\includegraphics[width=\columnwidth]{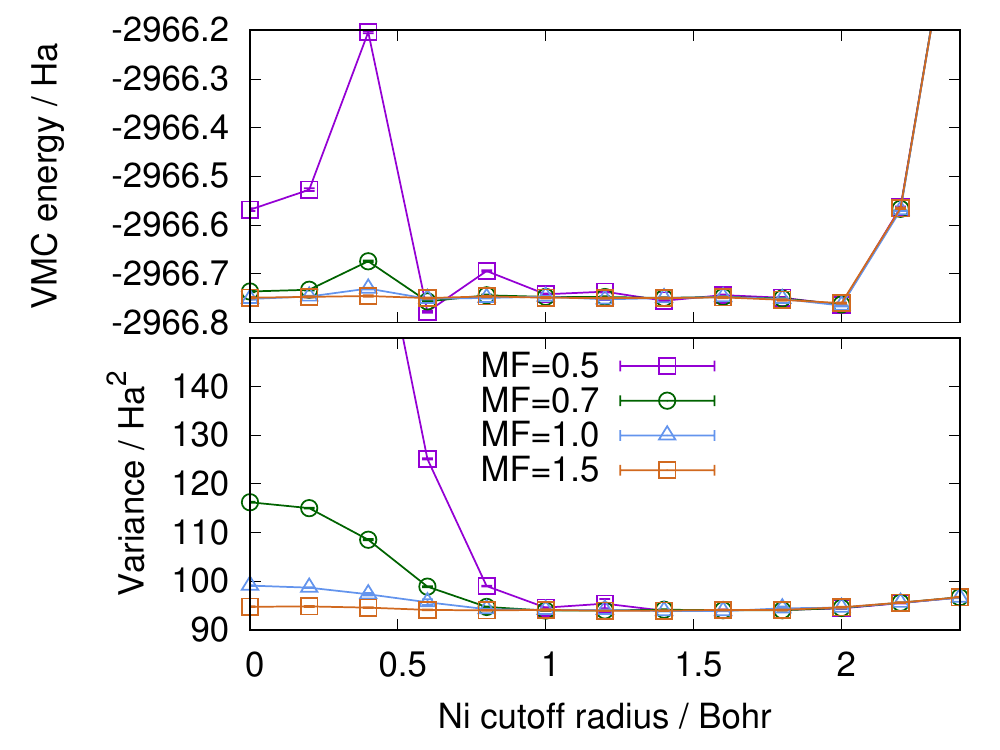}
\caption{VMC energy and variance with different cutoff radii applied on Ni atoms. O atoms always have a cutoff radius 1.4~bohr.}
\label{fig:Nicut}
\end{figure}

\begin{figure}[t]
\centering
\includegraphics[width=\columnwidth]{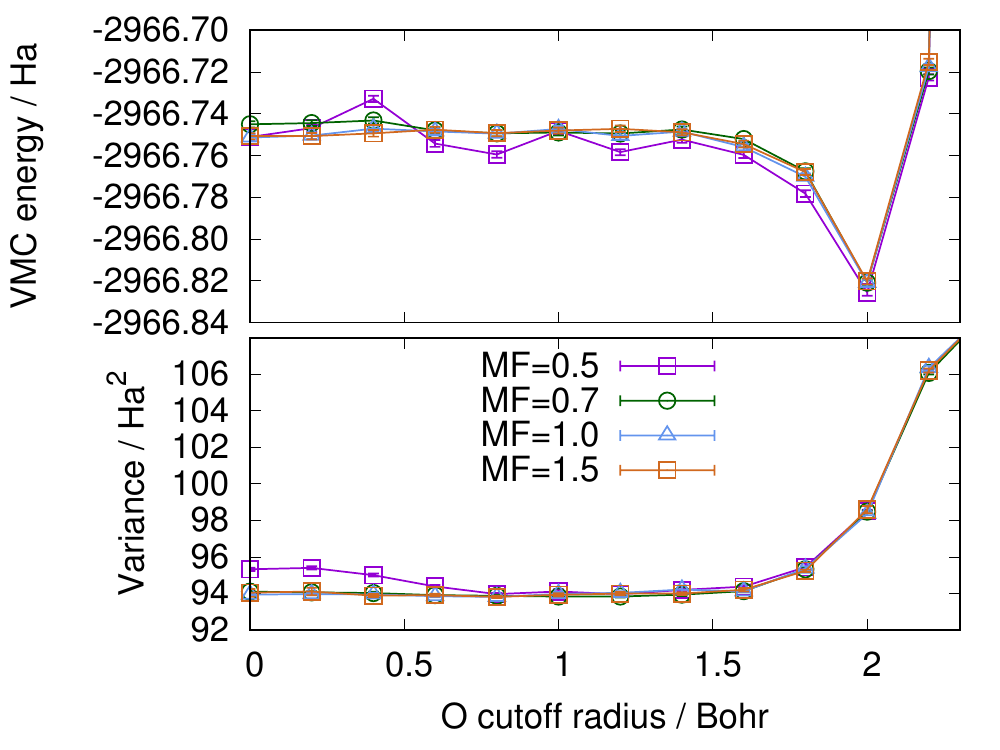}
\caption{VMC energy and variance with different cutoff radii applied on O atoms. Ni atoms always have a cutoff radius 1.4~bohr.}
\label{fig:Ocut}
\end{figure}

Then we study the impact varying $r_{\rm B/C}$ and using different mesh-factors. Fig.~\ref{fig:Nicut} shows a parametric scan for Ni atoms. These contribute the most significantly to the sensitivity of total energy due to their semi-core electrons. When there is only the B-spline representation ($r^{\rm Ni}_{\rm B/C}=0$), the energy is compromised and the variance ramps up quickly with small mesh-factors MF=0.7/0.5 as the wavefunction is inadequately described near the atomic cores.  When the atomic-like region A has a radius less than 1, the energy with small mesh-factors is not satisfactory although the variance drops quickly. There is even a non-monotonic behavior of the energy around $r^{\rm Ni}_{\rm B/C}=0.4$ caused by the large mismatch between atomic and B-spline representations. The smallest mesh-factor shows the largest accuracy loss though it has been alleviated somewhat by the smoothing. There is a safe region $r^{\rm Ni}_{\rm B/C} \in [1.0,1.8]$ where the energy remains highly accurate and the variance remains minimal. Beyond the safe region $r^{\rm Ni}_{\rm B/C} >1.8$, the variance starts to grow and the energy first decreases and then grows again for the same reason as the small $r^{\rm Ni}_{\rm B/C}$ region. By following the criteria that a good $r_{\rm B/C}$ should give both low energy and minimal variance, $r^{\rm O}_{\rm B/C} \in [0.8,1.4]$ is the safe region for O atoms based on Fig.~\ref{fig:Ocut}. We pick $r^{\rm Ni}_{\rm B/C}=1.4$ and $r^{\rm O}_{\rm B/C}=1.2$ for the remaining studies.

We notice that the Ni and O pseudopotentials have cutoff radii at 0.80 and 0.92\,bohr respectively which are similar to the safe regions of $r_{\rm B/C}$. This is not a coincidence: pseudopotentials are constructed to replace real deep attractive core potentials, and their cutoffs give the information on how large the atomic regions are. Our hybrid orbital representation is designed to represent atomic-like orbitals near core but is not efficient far from the ions. Thus we expect the radii where for the hybrid orbital regions to be similar to those of the pseudopotential core radii.

\subsection{DMC calculations}
DMC is the highest accuracy QMC method that is used routinely. In DMC, the nodal surface of the projected wavefunction is determined by the trial wavefunction. This must be represented in a sufficiently accurate basis set.  Besides representing the nodal surface, the trial wavefunction is also used for importance sampling. A poorer quality trial wavefunction will increase time step errors, and additional Monte Carlo steps will be required to obtain the same error due to increased variance.

\begin{figure}[t]
\centering
\includegraphics[width=\columnwidth]{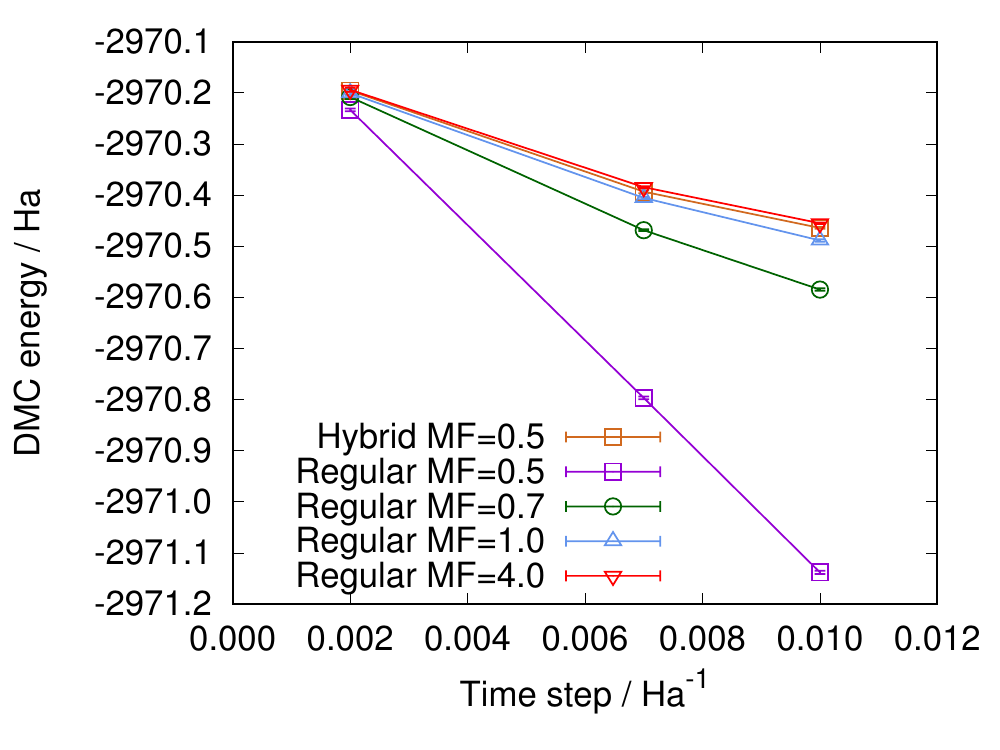}
\caption{DMC energy using different values of the time step. The hybrid orbital representation with MF=0.5 achieves the same accuracy as the regular representation with MF=4.0.}
\label{fig:tstep}
\end{figure}

\begin{figure}[t]
\centering
\includegraphics[width=\columnwidth]{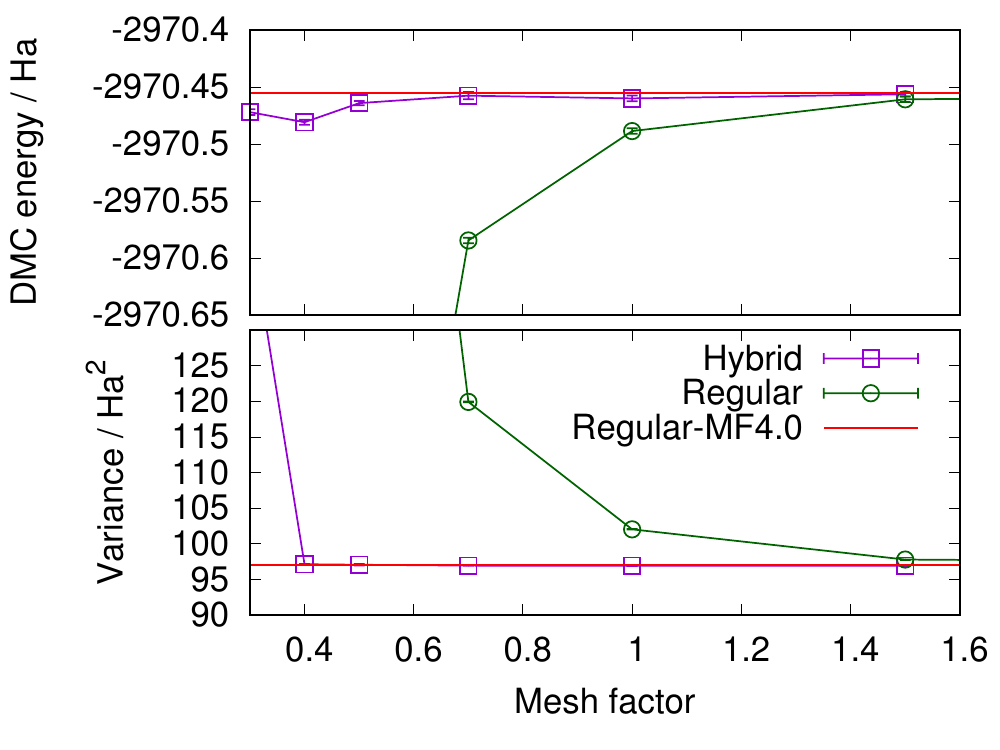}
\caption{DMC energy at different mesh-factors with regular or hybrid orbital representation. The time step $\tau$ is fixed at 0.01~Ha$^{-1}$.
 Hybrid orbital representation with MF=0.4 is still more accurate than regular representation with MF=1.0.}
\label{fig:mf}
\end{figure}

In Fig.~\ref{fig:tstep}, we show the energy for fixed time step and various choices of B-spline grid. The energy from DMC calculations with coarse B-spline grids departs significantly from a highly accurate but very memory intensive reference calculations with mesh factor equal to 4.0. When the hybrid orbital representation is used even with a very coarse grid (MF=0.5) for the interstitial region, the energy values remain on top of the ones calculated with only B-splines on a very dense grid. Interestingly, this wavefunction improves on the standard choice MF=1.0 by providing a smaller time step error and the same low variance as our much denser reference B-spline grid, in Fig.~\ref{fig:mf} while using around one eighth of the memory.  We also list the DMC energy measured with a $0.002$ time step in Tab.~\ref{tab:accuracy}.
With the hybrid orbital representation, it is therefore possible to use a substantially more affordable grid with MF=0.5 without compromising accuracy.

\begin{table}[t]
\centering
\begin{tabular}{|c|cc|cc|}
\hline
Rep. & MF & $\tau$ & Energy/Ha & Var./Ha$^2$ \\
\hline
hybrid & 0.3 & 0.002 & -2970.131(3) & 142.52(6) \\
hybrid & 0.4 & 0.002 & -2970.223(3) & 95.86(3) \\
hybrid & 0.5 & 0.002 & -2970.195(4) & 95.84(4) \\
hybrid & 0.7 & 0.002 & -2970.195(4) & 95.81(4) \\
hybrid & 1.0 & 0.002 & -2970.194(3) & 95.86(5) \\
hybrid & 1.5 & 0.002 & -2970.199(3) & 95.81(5) \\
\hline
regular & 0.5 & 0.002 & -2970.233(3) & 239.5(4) \\
regular & 0.7 & 0.002 & -2970.208(4) & 118.52(6) \\
regular & 1.0 & 0.002 & -2970.201(4) & 100.84(4) \\
regular & 1.5 & 0.002 & -2970.197(3) &  96.66(5) \\
regular & 4.0 & 0.002 & -2970.194(4) &  95.77(3) \\
\hline
\end{tabular}
\caption{DMC energy and variance at time step $\tau$=0.002. Hybrid orbital representation with MF$\ge$0.5 achieves the same accuracy as regular representation with MF=4.0.}
\label{tab:accuracy}
\end{table}

\subsection{Computational efficiency}
\begin{figure}[t]
\centering
\includegraphics[width=\columnwidth]{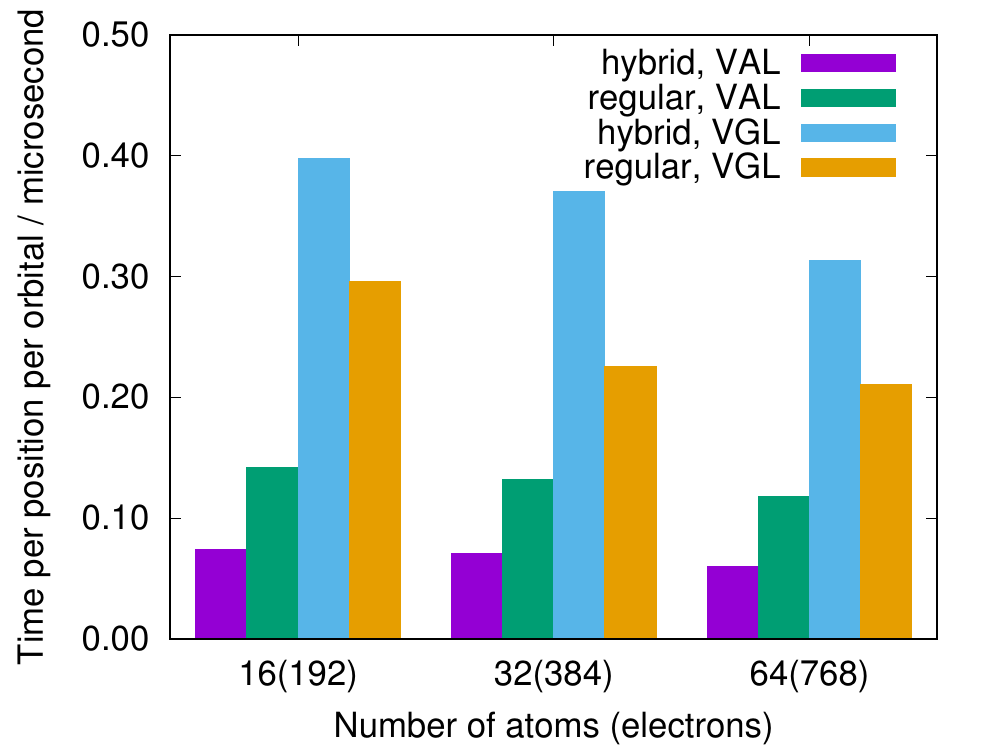}
\caption{Time spent on evaluating SPO values (VAL) or values with gradients and Laplacians (VGL) at a given electron position with hybrid and regular B-spline orbital representations.
The value is divided by the number of orbitals.
The measurement is performed on 8 Xeon Phi 7230 nodes with 64 walkers/threads per node.}
\label{fig:spo_time}
\end{figure}

The hybrid orbital representation is not just highly accurate and more compact, but it is also computationally very efficient for QMC.
Evaluation of regular B-spline orbitals is fast because of the explicitly localized computation and vectorization over all the orbitals.
Both characteristics are preserved in the hybrid orbital representation and thus it can perform well in both CPU SIMD and GPU SIMT architectures.
Table~\ref{tab:memory} compares the CPU time taken by DMC using both the highly tuned B-spline evaluation in QMCPACK\cite{ipdps,supercomputing} and this new hybrid orbital representation.
For comparable mesh sizes, the time to use the hybrid orbital representation is within a few percent of the simpler B-spline code.
The hybrid orbital representation actually requires more computation but for comparable accuracy, it has less pressure on cache and memory bandwidth thanks to its much smaller memory footprint.
Fig.~\ref{fig:spo_time} shows that the evaluation cost per orbital, theoretically constant, deceases gradually as the supercell size grows from 16 to 64 atoms.
It is due to the reducing fraction of constant cost from overhead when the total number of orbitals increases.
In all the supercell sizes, the VAL function of hybrid orbital representation has a better performance largely due to higher cache efficiency.
The VGL function with all problem sizes
takes about 50\% more time than regular B-splines, reflecting the increased computation of hybrid orbital representation.
This can be improved by using the cache blocking (tiling) technique~\cite{ipdps}.
The hybrid orbital representation with MF=0.5 needs only one eighth of the memory required by a regular representation with MF=1.0.
This not only enables much larger simulations to fit within the memory capacity constraints of current machines but also potentially allows faster computation
since more part of the orbitals are likely to fit higher in the memory hierarchy.

\begin{table}[t]
\centering
\begin{tabular}{|c|cc|cc|}
\hline
Rep. & MF & Mesh Size & Memory/MB & Time/sec\\
\hline
hybrid & 0.3 & 38$\times$22$\times$22 & 38 & 9.27(5) \\
hybrid & 0.4 & 50$\times$30$\times$30 & 73 & 9.26(6) \\
hybrid & 0.5 & 62$\times$36$\times$36 & 118 & 9.24(5) \\
hybrid & 0.7 & 82$\times$52$\times$52 & 312 & 9.27(6) \\
hybrid & 1.0 & 124$\times$72$\times$72 & 794 & 9.12(5) \\
hybrid & 1.5 & 192$\times$108$\times$108 & 2649 & 8.48(5) \\
\hline
regular & 0.5 & 62$\times$36$\times$36 & 108 & 9.57(10) \\
regular & 0.7 & 82$\times$52$\times$52 & 302 & 9.38(7) \\
regular & 1.0 & 124$\times$72$\times$72 & 784 & 9.02(4) \\
regular & 1.5 & 192$\times$108$\times$108 & 2639 & 7.93(6) \\
regular & 4.0 & 500$\times$288$\times$288 & 46795 & 8.76(3) \\
\hline
\end{tabular}
\caption{The memory usage (MB) of SPOs per spin and CPU time (seconds) for identical length blocks of DMC.
 Using the hybrid orbital representation significantly reduces the memory capacity demand without spending more time in computing.
The measurement is performed on 512 nodes of Xeon Phi 7230 with 128 walkers/threads per node.}
\label{tab:memory}
\end{table}

The hybrid orbital representation also lends itself to additional optimization.  Currently, evaluating the nonlocal part of the pseudopotential takes a significant proportion of time (30\% in our DMC test case) and it is mostly consumed by SPO evaluations.  With the hybrid orbital representation, the quadrature points for the required angular integration are all within the atomic region and a large amount of computation could be saved by evaluating the radial functions for all the quadrature points since they have the same distance from the core.

\section{Conclusions and Outlook}

We have developed a new representation of single particle orbitals for continuum quantum Monte Carlo simulations by hybridizing localized atomic basis sets and B-spline basis sets. An illustrative 32 atom NiO DMC simulation requires only one eighth of the memory of traditional B-splines, while demonstrating superior accuracy and maintaining high evaluation speed.

The hybrid basis set may be projected from plane wave or other basis sets.  At the moment, choosing the largest angular momentum in the atomic basis sets and deducing the optimal spherical radii are additional required steps to properly use the hybrid orbital representation.  However, we have found these choices to be nearly independent of the particular system other than the pseudopotentials used for each ion.  This complication could be resolved in the future by including standard choices for the cutoff radius and largest angular momentum channel as part of the pseudopotential data files intended for QMC calculations.

The hybrid orbital representation is also well suited to exploit current trends in the evolution of high performance computing hardware.  Computational resources have become increasingly heterogeneous with hierarchical memory subsystems. The small data set for the localized atomic basis sets can be placed in fast but limited capacity memory while the large data set for the B-spline basis sets can be left in slower but high capacity memory. In this way, the performance penalty from the memory hierarchy is reduced.

As QMC simulations mature from simple homogeneous systems to complex and inhomogeneous system containing heavier elements and more electrons, we anticipate our new approach will enable significantly more challenging problems to be tackled on current and upcoming generations of supercomputers.

\section*{Acknowledgments}
This work was supported by the U.S. Department of Energy, Office of Science, Basic Energy
Sciences, Materials Sciences and Engineering Division, as part of the Computational Materials
Sciences Program and Center for Predictive Simulation of Functional Materials. An award of computer time was provided by the Innovative and Novel Computational Impact on Theory and Experiment (INCITE) program. This research used resources of the Argonne Leadership Computing Facility, which is a DOE Office of Science User Facility supported under Contract DE-AC02-06CH11357. Sandia National Laboratories is a multimission laboratory managed and operated by National Technology and Engineering Solutions of Sandia, LLC, a wholly owned subsidiary of Honeywell International Inc., for the U.S. Department of Energy’s National Nuclear Security Administration under contract DE-NA0003525. The views expressed in the article do not necessarily represent the views of the U.S. Department of Energy or the United States Government.

\appendix

\section{Construction of radial functions}
\label{app:parallel}
In the hybrid orbital representation, the radial functions of atomic orbitals are transformed
by projecting the original orbitals onto a basis of real spherical harmonics at every atomic center.
The input orbitals from DFT calculations are expressed in PW basis as
\begin{equation}
\phi^n(\mathbf{r}) = \sum_{\mathbf{G}} c^n_{\mathbf G} e^{-i \mathbf{G} \cdot \mathbf{r}} \quad .
\end{equation}
Since each PW can be expressed with complex spherical harmonics $\bar{Y}_{l,m}$ using the spherical harmonic addition theorem,
we first expand the orbitals on $\bar{Y}_{l,m}$ and then change the basis set to the real spherical harmonics $Y_{l,m}$.
Finally, the radial functions can be simply computed as
\begin{equation}
u^{n,\mathbf{R}}_{l,m}(r)  = 4\pi i^l \sum_{\mathbf{G}} c^n_{\mathbf G} e^{i \mathbf{G} \cdot \mathbf{R}} j_l(G r) Y_{l,m}(\hat{G})
\end{equation}
where $j_l$ is the spherical Bessel function and $e^{i \mathbf{G} \cdot \mathbf{R}}$ is the phase shift for the centers.
Their computation dominates the extremely expensive construction cost
which grows as $N_{\rm orbs} \times N_{\rm PW} \times N_{\rm atoms} \times C_{\rm grid} \times (l_{\rm max}+1)^2$.

To obtain radial functions rapidly as possible, we parallelize the computation over MPI ranks (nodes) and OpenMP threads (cores).
QMCPACK already distributes the orbitals over MPI ranks during the initialization and thus it is also adopted for our needs.
The summation over the PW basis is then distributed among the ranks serving the same band.
Within a node, the subset of the PW basis set is chunked into tiles and then distributed among all the cores via OpenMP threads.
The tiling implementation not only facilitates the threading but also provides cache blocking.
We also fully vectorized the multiplication and accumulation in the innermost loops.
With all the parallelization and optimization above, the heavy initialization of the hybrid orbital representation becomes a small cost for almost all current applications.
For the 32 atom NiO supercell calculation on 8 nodes of Xeon Phi 7230,
the full initialization of regular B-spline representation from the 200\,Ha plane-wave cutoff orbitals takes 5.9 seconds.
Using hybrid orbital representation only adds 2.0 seconds.

We also experimented with computing radial functions by projecting real space wavefunction on spherical harmonics.
This route is appealing because only a finite subset of the real space grid is necessary for each projection.
The projection is conducted by using quadrature integration and the needed real-space
orbital values are evaluated via B-spline interpolation on a very dense grid
which can be constructed using the fast Fourier transform. Unfortunately this route did not
provide sufficient numerical accuracy due to significant error from quadrature integration at large distances,
even if the orbital values were computed directly from PWs and higher order quadrature rules were used.
The error is small very near the core ($<0.2$\,a.u.) but grows beyond the necessary accuracy
at $\approx 1.0$\,a.u. where most of the spherical region radius lives.
For this reason, this route is not considered practical.

\section{Computing the gradient and Laplacian}
\label{app:gradlapl}
To calculate the kinetic part of the local energy, the gradients and Laplacian of SPOs need to be computed at given electronic positions.
In our scheme, we break the computation of $\phi(\mathbf{r})$ into separate parts for the $Y_{l,m}(\hat{r})$ and $u(r)$.
In the $Y_{l,m}$ part, the gradients and Laplacian are computed from $r^l Y_{l,m}$
instead of $Y_{l,m}$ in order to exploit a nice property of $r^l Y_{l,m}$
\begin{align}
  {\mathbf G} & = \nabla ( r^l Y_{l,m} ) = l \hat{r} r^{l-1} Y_{l,m} + r^l \nabla Y_{l,m} \\
  0 & = \lapl ( r^l Y_{l,m} ) \\
    & = l r^{l-2} Y_{l,m} + l r^{-1} \hat{r}\cdot {\mathbf G} + r^l \lapl Y_{l,m}
\end{align}
The contribution from the $u(r)$ part is directly computed as
\begin{align}
\nabla u(r) & = g_u \hat{r} & g_u & = \frac{\partial u (r)}{\partial r} \\
\lapl u(r) & = l_u + g_u \frac{2}{r} & l_u & = \frac{\partial^2 u (r)}{\partial r^2}
\end{align}
The combined gradients and Laplacian of $\phi(\mathbf{r})$ are thus computed as
\begin{align}
\nabla (u(r)Y_{l,m}) & = g_u \hat{r} Y_{l,m} + \frac{u(r)}{r} ( {\mathbf G} r^{1-l} - l \hat{r} Y_{l,m} ) \\
\lapl (u(r)Y_{l,m}) & = \Big(l_u + g_u \frac{2}{r} \Big) Y_{l,m} \\
                   & + \frac{g_u}{r} (\hat{r}\cdot {\mathbf G} r^{1-l} - lY_{l,m}) \\
                   & - \frac{u(r)}{r^2} l (\hat{r}\cdot {\mathbf G} r^{1-l} + Y_{l,m})
\end{align}
At small $r$, the computation is replaced with an asymptotic formula to avoid the divergence of $1/r$;


\begin{thebibliography}{49}
\expandafter\ifx\csname natexlab\endcsname\relax\def\natexlab#1{#1}\fi
\expandafter\ifx\csname bibnamefont\endcsname\relax
  \def\bibnamefont#1{#1}\fi
\expandafter\ifx\csname bibfnamefont\endcsname\relax
  \def\bibfnamefont#1{#1}\fi
\expandafter\ifx\csname citenamefont\endcsname\relax
  \def\citenamefont#1{#1}\fi
\expandafter\ifx\csname url\endcsname\relax
  \def\url#1{\texttt{#1}}\fi
\expandafter\ifx\csname urlprefix\endcsname\relax\def\urlprefix{URL }\fi
\providecommand{\bibinfo}[2]{#2}
\providecommand{\eprint}[2][]{\url{#2}}

\bibitem[{\citenamefont{Foulkes et~al.}(2001)\citenamefont{Foulkes, Mitas,
  Needs, and Rajagopal}}]{rev:qmcsolids}
\bibinfo{author}{\bibfnamefont{W.~M.~C.} \bibnamefont{Foulkes}},
  \bibinfo{author}{\bibfnamefont{L.}~\bibnamefont{Mitas}},
  \bibinfo{author}{\bibfnamefont{R.~J.} \bibnamefont{Needs}}, \bibnamefont{and}
  \bibinfo{author}{\bibfnamefont{G.}~\bibnamefont{Rajagopal}},
  \bibinfo{journal}{Reviews of Modern Physics} \textbf{\bibinfo{volume}{73}},
  \bibinfo{pages}{33} (\bibinfo{year}{2001}).

\bibitem[{\citenamefont{Needs et~al.}(2010)\citenamefont{Needs, Towler,
  Drummond, and {L{\'{o}}pez R{\'{i}}os}}}]{Needs2010}
\bibinfo{author}{\bibfnamefont{R.~J.} \bibnamefont{Needs}},
  \bibinfo{author}{\bibfnamefont{M.~D.} \bibnamefont{Towler}},
  \bibinfo{author}{\bibfnamefont{N.~D.} \bibnamefont{Drummond}},
  \bibnamefont{and}
  \bibinfo{author}{\bibfnamefont{P.}~\bibnamefont{{L{\'{o}}pez R{\'{i}}os}}},
  \bibinfo{journal}{Journal of Physics Condensed Matter}
  \textbf{\bibinfo{volume}{22}}, \bibinfo{pages}{023201}
  (\bibinfo{year}{2010}).

\bibitem[{\citenamefont{Benali et~al.}(2014)\citenamefont{Benali, Shulenburger,
  Romero, Kim, and von Lilienfeld}}]{Benali2014}
\bibinfo{author}{\bibfnamefont{A.}~\bibnamefont{Benali}},
  \bibinfo{author}{\bibfnamefont{L.}~\bibnamefont{Shulenburger}},
  \bibinfo{author}{\bibfnamefont{N.~A.} \bibnamefont{Romero}},
  \bibinfo{author}{\bibfnamefont{J.}~\bibnamefont{Kim}}, \bibnamefont{and}
  \bibinfo{author}{\bibfnamefont{O.~A.} \bibnamefont{von Lilienfeld}},
  \bibinfo{journal}{Journal of Chemical Theory and Computation}
  \textbf{\bibinfo{volume}{10}}, \bibinfo{pages}{3417} (\bibinfo{year}{2014}).

\bibitem[{\citenamefont{Krongchon et~al.}(2017)\citenamefont{Krongchon,
  Busemeyer, and Wagner}}]{Krongchon2017}
\bibinfo{author}{\bibfnamefont{K.}~\bibnamefont{Krongchon}},
  \bibinfo{author}{\bibfnamefont{B.}~\bibnamefont{Busemeyer}},
  \bibnamefont{and} \bibinfo{author}{\bibfnamefont{L.~K.}
  \bibnamefont{Wagner}}, \bibinfo{journal}{The Journal of Chemical Physics}
  \textbf{\bibinfo{volume}{146}}, \bibinfo{pages}{124129}
  (\bibinfo{year}{2017}).

\bibitem[{\citenamefont{Zen et~al.}(2018)\citenamefont{Zen, Brandenburg,
  Klime{\v{s}}, Tkatchenko, Alf{\`{e}}, and Michaelides}}]{Zen2018}
\bibinfo{author}{\bibfnamefont{A.}~\bibnamefont{Zen}},
  \bibinfo{author}{\bibfnamefont{J.~G.} \bibnamefont{Brandenburg}},
  \bibinfo{author}{\bibfnamefont{J.}~\bibnamefont{Klime{\v{s}}}},
  \bibinfo{author}{\bibfnamefont{A.}~\bibnamefont{Tkatchenko}},
  \bibinfo{author}{\bibfnamefont{D.}~\bibnamefont{Alf{\`{e}}}},
  \bibnamefont{and}
  \bibinfo{author}{\bibfnamefont{A.}~\bibnamefont{Michaelides}},
  \bibinfo{journal}{Proceedings of the National Academy of Sciences}
  \textbf{\bibinfo{volume}{115}}, \bibinfo{pages}{1724} (\bibinfo{year}{2018}).

\bibitem[{\citenamefont{Zen et~al.}(2016)\citenamefont{Zen, Sorella, Gillan,
  Michaelides, and Alf{\`e}}}]{Zen2016}
\bibinfo{author}{\bibfnamefont{A.}~\bibnamefont{Zen}},
  \bibinfo{author}{\bibfnamefont{S.}~\bibnamefont{Sorella}},
  \bibinfo{author}{\bibfnamefont{M.~J.} \bibnamefont{Gillan}},
  \bibinfo{author}{\bibfnamefont{A.}~\bibnamefont{Michaelides}},
  \bibnamefont{and} \bibinfo{author}{\bibfnamefont{D.}~\bibnamefont{Alf{\`e}}},
  \bibinfo{journal}{Physical Review B - Condensed Matter and Materials Physics}
  \textbf{\bibinfo{volume}{93}}, \bibinfo{pages}{1} (\bibinfo{year}{2016}).

\bibitem[{\citenamefont{Mouhat et~al.}(2017)\citenamefont{Mouhat, Sorella,
  Vuilleumier, Saitta, and Casula}}]{Mouhat2017}
\bibinfo{author}{\bibfnamefont{F.}~\bibnamefont{Mouhat}},
  \bibinfo{author}{\bibfnamefont{S.}~\bibnamefont{Sorella}},
  \bibinfo{author}{\bibfnamefont{R.}~\bibnamefont{Vuilleumier}},
  \bibinfo{author}{\bibfnamefont{A.~M.} \bibnamefont{Saitta}},
  \bibnamefont{and} \bibinfo{author}{\bibfnamefont{M.}~\bibnamefont{Casula}},
  \bibinfo{journal}{Journal of Chemical Theory and Computation}
  \textbf{\bibinfo{volume}{13}}, \bibinfo{pages}{2400} (\bibinfo{year}{2017}).

\bibitem[{\citenamefont{Motta et~al.}(2017)\citenamefont{Motta, Ceperley, Chan,
  Gomez, Gull, Guo, Jim{\'{e}}nez-Hoyos, Lan, Li, Ma et~al.}}]{Motta2017}
\bibinfo{author}{\bibfnamefont{M.}~\bibnamefont{Motta}},
  \bibinfo{author}{\bibfnamefont{D.~M.} \bibnamefont{Ceperley}},
  \bibinfo{author}{\bibfnamefont{G.~K.-L.} \bibnamefont{Chan}},
  \bibinfo{author}{\bibfnamefont{J.~A.} \bibnamefont{Gomez}},
  \bibinfo{author}{\bibfnamefont{E.}~\bibnamefont{Gull}},
  \bibinfo{author}{\bibfnamefont{S.}~\bibnamefont{Guo}},
  \bibinfo{author}{\bibfnamefont{C.~A.} \bibnamefont{Jim{\'{e}}nez-Hoyos}},
  \bibinfo{author}{\bibfnamefont{T.~N.} \bibnamefont{Lan}},
  \bibinfo{author}{\bibfnamefont{J.}~\bibnamefont{Li}},
  \bibinfo{author}{\bibfnamefont{F.}~\bibnamefont{Ma}}, \bibnamefont{et~al.},
  \bibinfo{journal}{Physical Review X} \textbf{\bibinfo{volume}{7}},
  \bibinfo{pages}{031059} (\bibinfo{year}{2017}).

\bibitem[{\citenamefont{Fumanal et~al.}(2016)\citenamefont{Fumanal, Wagner,
  Sanvito, and Droghetti}}]{Fumanal2016}
\bibinfo{author}{\bibfnamefont{M.}~\bibnamefont{Fumanal}},
  \bibinfo{author}{\bibfnamefont{L.~K.} \bibnamefont{Wagner}},
  \bibinfo{author}{\bibfnamefont{S.}~\bibnamefont{Sanvito}}, \bibnamefont{and}
  \bibinfo{author}{\bibfnamefont{A.}~\bibnamefont{Droghetti}},
  \bibinfo{journal}{Journal of Chemical Theory and Computation}
  \textbf{\bibinfo{volume}{12}}, \bibinfo{pages}{4233} (\bibinfo{year}{2016}).

\bibitem[{\citenamefont{Chen et~al.}(2016)\citenamefont{Chen, Zen, Brandenburg,
  Alf{\`{e}}, and Michaelides}}]{Chen2016}
\bibinfo{author}{\bibfnamefont{J.}~\bibnamefont{Chen}},
  \bibinfo{author}{\bibfnamefont{A.}~\bibnamefont{Zen}},
  \bibinfo{author}{\bibfnamefont{J.~G.} \bibnamefont{Brandenburg}},
  \bibinfo{author}{\bibfnamefont{D.}~\bibnamefont{Alf{\`{e}}}},
  \bibnamefont{and}
  \bibinfo{author}{\bibfnamefont{A.}~\bibnamefont{Michaelides}},
  \bibinfo{journal}{Physical Review B} \textbf{\bibinfo{volume}{94}},
  \bibinfo{pages}{220102} (\bibinfo{year}{2016}).

\bibitem[{\citenamefont{Luo et~al.}(2016)\citenamefont{Luo, Benali,
  Shulenburger, Krogel, Heinonen, and Kent}}]{Luo2016}
\bibinfo{author}{\bibfnamefont{Y.}~\bibnamefont{Luo}},
  \bibinfo{author}{\bibfnamefont{A.}~\bibnamefont{Benali}},
  \bibinfo{author}{\bibfnamefont{L.}~\bibnamefont{Shulenburger}},
  \bibinfo{author}{\bibfnamefont{J.~T.} \bibnamefont{Krogel}},
  \bibinfo{author}{\bibfnamefont{O.}~\bibnamefont{Heinonen}}, \bibnamefont{and}
  \bibinfo{author}{\bibfnamefont{P.~R.~C.} \bibnamefont{Kent}},
  \bibinfo{journal}{New Journal of Physics} \textbf{\bibinfo{volume}{18}},
  \bibinfo{pages}{113049} (\bibinfo{year}{2016}).

\bibitem[{\citenamefont{Tsatsoulis et~al.}(2017)\citenamefont{Tsatsoulis,
  Hummel, Usvyat, Sch{\"{u}}tz, Booth, Binnie, Gillan, Alf{\`{e}}, Michaelides,
  and Gr{\"{u}}neis}}]{Tsatsoulis2017}
\bibinfo{author}{\bibfnamefont{T.}~\bibnamefont{Tsatsoulis}},
  \bibinfo{author}{\bibfnamefont{F.}~\bibnamefont{Hummel}},
  \bibinfo{author}{\bibfnamefont{D.}~\bibnamefont{Usvyat}},
  \bibinfo{author}{\bibfnamefont{M.}~\bibnamefont{Sch{\"{u}}tz}},
  \bibinfo{author}{\bibfnamefont{G.~H.} \bibnamefont{Booth}},
  \bibinfo{author}{\bibfnamefont{S.~S.} \bibnamefont{Binnie}},
  \bibinfo{author}{\bibfnamefont{M.~J.} \bibnamefont{Gillan}},
  \bibinfo{author}{\bibfnamefont{D.}~\bibnamefont{Alf{\`{e}}}},
  \bibinfo{author}{\bibfnamefont{A.}~\bibnamefont{Michaelides}},
  \bibnamefont{and}
  \bibinfo{author}{\bibfnamefont{A.}~\bibnamefont{Gr{\"{u}}neis}},
  \textbf{\bibinfo{volume}{204108}} (\bibinfo{year}{2017}).

\bibitem[{\citenamefont{Azadi et~al.}(2017)\citenamefont{Azadi, Drummond, and
  Foulkes}}]{Azadi2017}
\bibinfo{author}{\bibfnamefont{S.}~\bibnamefont{Azadi}},
  \bibinfo{author}{\bibfnamefont{N.~D.} \bibnamefont{Drummond}},
  \bibnamefont{and} \bibinfo{author}{\bibfnamefont{W.~M.~C.}
  \bibnamefont{Foulkes}}, \bibinfo{journal}{Physical Review B}
  \textbf{\bibinfo{volume}{95}}, \bibinfo{pages}{035142}
  (\bibinfo{year}{2017}).

\bibitem[{\citenamefont{Trail et~al.}(2017)\citenamefont{Trail, Monserrat,
  {L{\'{o}}pez R{\'{i}}os}, Maezono, and Needs}}]{Trail2017}
\bibinfo{author}{\bibfnamefont{J.}~\bibnamefont{Trail}},
  \bibinfo{author}{\bibfnamefont{B.}~\bibnamefont{Monserrat}},
  \bibinfo{author}{\bibfnamefont{P.}~\bibnamefont{{L{\'{o}}pez R{\'{i}}os}}},
  \bibinfo{author}{\bibfnamefont{R.}~\bibnamefont{Maezono}}, \bibnamefont{and}
  \bibinfo{author}{\bibfnamefont{R.~J.} \bibnamefont{Needs}},
  \bibinfo{journal}{Physical Review B} \textbf{\bibinfo{volume}{95}},
  \bibinfo{pages}{121108} (\bibinfo{year}{2017}).

\bibitem[{\citenamefont{Varsano et~al.}(2017)\citenamefont{Varsano, Sorella,
  Sangalli, Barborini, Corni, Molinari, and Rontani}}]{Varsano2017}
\bibinfo{author}{\bibfnamefont{D.}~\bibnamefont{Varsano}},
  \bibinfo{author}{\bibfnamefont{S.}~\bibnamefont{Sorella}},
  \bibinfo{author}{\bibfnamefont{D.}~\bibnamefont{Sangalli}},
  \bibinfo{author}{\bibfnamefont{M.}~\bibnamefont{Barborini}},
  \bibinfo{author}{\bibfnamefont{S.}~\bibnamefont{Corni}},
  \bibinfo{author}{\bibfnamefont{E.}~\bibnamefont{Molinari}}, \bibnamefont{and}
  \bibinfo{author}{\bibfnamefont{M.}~\bibnamefont{Rontani}},
  \bibinfo{journal}{Nature Communications} \textbf{\bibinfo{volume}{8}},
  \bibinfo{pages}{1461} (\bibinfo{year}{2017}).

\bibitem[{\citenamefont{Mazzola et~al.}(2018)\citenamefont{Mazzola, Helled, and
  Sorella}}]{Mazzola2018}
\bibinfo{author}{\bibfnamefont{G.}~\bibnamefont{Mazzola}},
  \bibinfo{author}{\bibfnamefont{R.}~\bibnamefont{Helled}}, \bibnamefont{and}
  \bibinfo{author}{\bibfnamefont{S.}~\bibnamefont{Sorella}},
  \bibinfo{journal}{Physical Review Letters} \textbf{\bibinfo{volume}{120}},
  \bibinfo{pages}{025701} (\bibinfo{year}{2018}).

\bibitem[{\citenamefont{Shin et~al.}(2017)\citenamefont{Shin, Luo, Ganesh,
  Balachandran, Krogel, Kent, Benali, and Heinonen}}]{Shin2017}
\bibinfo{author}{\bibfnamefont{H.}~\bibnamefont{Shin}},
  \bibinfo{author}{\bibfnamefont{Y.}~\bibnamefont{Luo}},
  \bibinfo{author}{\bibfnamefont{P.}~\bibnamefont{Ganesh}},
  \bibinfo{author}{\bibfnamefont{J.}~\bibnamefont{Balachandran}},
  \bibinfo{author}{\bibfnamefont{J.~T.} \bibnamefont{Krogel}},
  \bibinfo{author}{\bibfnamefont{P.~R.~C.} \bibnamefont{Kent}},
  \bibinfo{author}{\bibfnamefont{A.}~\bibnamefont{Benali}}, \bibnamefont{and}
  \bibinfo{author}{\bibfnamefont{O.}~\bibnamefont{Heinonen}},
  \bibinfo{journal}{Physical Review Materials} \textbf{\bibinfo{volume}{1}},
  \bibinfo{pages}{073603} (\bibinfo{year}{2017}).

\bibitem[{\citenamefont{Kyl{\"{a}}np{\"{a}}{\"{a}}
  et~al.}(2017)\citenamefont{Kyl{\"{a}}np{\"{a}}{\"{a}}, Balachandran, Ganesh,
  Heinonen, Kent, and Krogel}}]{Kylanpaa2017}
\bibinfo{author}{\bibfnamefont{I.}~\bibnamefont{Kyl{\"{a}}np{\"{a}}{\"{a}}}},
  \bibinfo{author}{\bibfnamefont{J.}~\bibnamefont{Balachandran}},
  \bibinfo{author}{\bibfnamefont{P.}~\bibnamefont{Ganesh}},
  \bibinfo{author}{\bibfnamefont{O.}~\bibnamefont{Heinonen}},
  \bibinfo{author}{\bibfnamefont{P.~R.~C.} \bibnamefont{Kent}},
  \bibnamefont{and} \bibinfo{author}{\bibfnamefont{J.~T.}
  \bibnamefont{Krogel}}, \bibinfo{journal}{Physical Review Materials}
  \textbf{\bibinfo{volume}{1}}, \bibinfo{pages}{065408} (\bibinfo{year}{2017}).

\bibitem[{\citenamefont{Dzubak et~al.}(2017{\natexlab{a}})\citenamefont{Dzubak,
  Mitra, Chance, Kuhn, Jellison, Sefat, Krogel, and Reboredo}}]{Dzubak2017}
\bibinfo{author}{\bibfnamefont{A.~L.} \bibnamefont{Dzubak}},
  \bibinfo{author}{\bibfnamefont{C.}~\bibnamefont{Mitra}},
  \bibinfo{author}{\bibfnamefont{M.}~\bibnamefont{Chance}},
  \bibinfo{author}{\bibfnamefont{S.}~\bibnamefont{Kuhn}},
  \bibinfo{author}{\bibfnamefont{G.~E.} \bibnamefont{Jellison}},
  \bibinfo{author}{\bibfnamefont{A.~S.} \bibnamefont{Sefat}},
  \bibinfo{author}{\bibfnamefont{J.~T.} \bibnamefont{Krogel}},
  \bibnamefont{and} \bibinfo{author}{\bibfnamefont{F.~A.}
  \bibnamefont{Reboredo}}, \bibinfo{journal}{The Journal of Chemical Physics}
  \textbf{\bibinfo{volume}{147}}, \bibinfo{pages}{174703}
  (\bibinfo{year}{2017}{\natexlab{a}}).

\bibitem[{\citenamefont{Santana et~al.}(2017)\citenamefont{Santana, Krogel,
  Kent, and Reboredo}}]{Santana2017}
\bibinfo{author}{\bibfnamefont{J.~A.} \bibnamefont{Santana}},
  \bibinfo{author}{\bibfnamefont{J.~T.} \bibnamefont{Krogel}},
  \bibinfo{author}{\bibfnamefont{P.~R.~C.} \bibnamefont{Kent}},
  \bibnamefont{and} \bibinfo{author}{\bibfnamefont{F.~A.}
  \bibnamefont{Reboredo}}, \bibinfo{journal}{The Journal of Chemical Physics}
  \textbf{\bibinfo{volume}{147}}, \bibinfo{pages}{034701}
  (\bibinfo{year}{2017}).

\bibitem[{\citenamefont{Yu et~al.}(2017)\citenamefont{Yu, Wagner, and
  Ertekin}}]{Yu2017}
\bibinfo{author}{\bibfnamefont{J.}~\bibnamefont{Yu}},
  \bibinfo{author}{\bibfnamefont{L.~K.} \bibnamefont{Wagner}},
  \bibnamefont{and} \bibinfo{author}{\bibfnamefont{E.}~\bibnamefont{Ertekin}},
  \bibinfo{journal}{Physical Review B} \textbf{\bibinfo{volume}{95}},
  \bibinfo{pages}{075209} (\bibinfo{year}{2017}).

\bibitem[{\citenamefont{Saritas et~al.}(2017)\citenamefont{Saritas, Mueller,
  Wagner, and Grossman}}]{Saritas2017}
\bibinfo{author}{\bibfnamefont{K.}~\bibnamefont{Saritas}},
  \bibinfo{author}{\bibfnamefont{T.}~\bibnamefont{Mueller}},
  \bibinfo{author}{\bibfnamefont{L.}~\bibnamefont{Wagner}}, \bibnamefont{and}
  \bibinfo{author}{\bibfnamefont{J.~C.} \bibnamefont{Grossman}},
  \bibinfo{journal}{Journal of Chemical Theory and Computation}
  \textbf{\bibinfo{volume}{13}}, \bibinfo{pages}{1943} (\bibinfo{year}{2017}).

\bibitem[{\citenamefont{Dubeck{\'{y}} et~al.}(2016)\citenamefont{Dubeck{\'{y}},
  Mitas, and Jure{\v{c}}ka}}]{Dubecky2016}
\bibinfo{author}{\bibfnamefont{M.}~\bibnamefont{Dubeck{\'{y}}}},
  \bibinfo{author}{\bibfnamefont{L.}~\bibnamefont{Mitas}}, \bibnamefont{and}
  \bibinfo{author}{\bibfnamefont{P.}~\bibnamefont{Jure{\v{c}}ka}},
  \bibinfo{journal}{Chemical Reviews} \textbf{\bibinfo{volume}{116}},
  \bibinfo{pages}{5188} (\bibinfo{year}{2016}).

\bibitem[{\citenamefont{Wagner and Ceperley}(2016)}]{Wagner2016}
\bibinfo{author}{\bibfnamefont{L.~K.} \bibnamefont{Wagner}} \bibnamefont{and}
  \bibinfo{author}{\bibfnamefont{D.~M.} \bibnamefont{Ceperley}},
  \bibinfo{journal}{Reports on Progress in Physics}
  \textbf{\bibinfo{volume}{79}} (\bibinfo{year}{2016}).

\bibitem[{\citenamefont{McMillan}(1965)}]{McMillan1965}
\bibinfo{author}{\bibfnamefont{W.~L.} \bibnamefont{McMillan}},
  \bibinfo{journal}{Physical Review} \textbf{\bibinfo{volume}{138}},
  \bibinfo{pages}{A442} (\bibinfo{year}{1965}).

\bibitem[{\citenamefont{Grimm and Storer}(1971)}]{Grimm1971}
\bibinfo{author}{\bibfnamefont{R.}~\bibnamefont{Grimm}} \bibnamefont{and}
  \bibinfo{author}{\bibfnamefont{R.}~\bibnamefont{Storer}},
  \bibinfo{journal}{Journal of Computational Physics}
  \textbf{\bibinfo{volume}{7}}, \bibinfo{pages}{134} (\bibinfo{year}{1971}).

\bibitem[{\citenamefont{Kalos et~al.}(1974)\citenamefont{Kalos, Levesque, and
  Verlet}}]{Kalos1974}
\bibinfo{author}{\bibfnamefont{M.~H.} \bibnamefont{Kalos}},
  \bibinfo{author}{\bibfnamefont{D.}~\bibnamefont{Levesque}}, \bibnamefont{and}
  \bibinfo{author}{\bibfnamefont{L.}~\bibnamefont{Verlet}},
  \bibinfo{journal}{Physical Review A} \textbf{\bibinfo{volume}{9}},
  \bibinfo{pages}{2178} (\bibinfo{year}{1974}).

\bibitem[{\citenamefont{Hood et~al.}(2012)\citenamefont{Hood, Kent, and
  Reboredo}}]{Hood2012}
\bibinfo{author}{\bibfnamefont{R.~Q.} \bibnamefont{Hood}},
  \bibinfo{author}{\bibfnamefont{P.~R.~C.} \bibnamefont{Kent}},
  \bibnamefont{and} \bibinfo{author}{\bibfnamefont{F.~A.}
  \bibnamefont{Reboredo}}, \bibinfo{journal}{Physical Review B}
  \textbf{\bibinfo{volume}{85}}, \bibinfo{pages}{134109}
  (\bibinfo{year}{2012}).

\bibitem[{\citenamefont{Alf{\`{e}} and Gillan}(2004{\natexlab{a}})}]{Alfe2004}
\bibinfo{author}{\bibfnamefont{D.}~\bibnamefont{Alf{\`{e}}}} \bibnamefont{and}
  \bibinfo{author}{\bibfnamefont{M.~J.} \bibnamefont{Gillan}},
  \textbf{\bibinfo{volume}{161101}}, \bibinfo{pages}{4}
  (\bibinfo{year}{2004}{\natexlab{a}}).

\bibitem[{\citenamefont{Parker et~al.}(2015)\citenamefont{Parker, Umrigar,
  Alfè, Petruzielo, Hennig, and Wilkins}}]{parker2015}
\bibinfo{author}{\bibfnamefont{W.~D.} \bibnamefont{Parker}},
  \bibinfo{author}{\bibfnamefont{C.}~\bibnamefont{Umrigar}},
  \bibinfo{author}{\bibfnamefont{D.}~\bibnamefont{Alfè}},
  \bibinfo{author}{\bibfnamefont{F.}~\bibnamefont{Petruzielo}},
  \bibinfo{author}{\bibfnamefont{R.~G.} \bibnamefont{Hennig}},
  \bibnamefont{and} \bibinfo{author}{\bibfnamefont{J.~W.}
  \bibnamefont{Wilkins}}, \bibinfo{journal}{Journal of Computational Physics}
  \textbf{\bibinfo{volume}{287}}, \bibinfo{pages}{77 } (\bibinfo{year}{2015}).

\bibitem[{\citenamefont{Reboredo and Williamson}(2005)}]{Reboredo2005}
\bibinfo{author}{\bibfnamefont{F.~A.} \bibnamefont{Reboredo}} \bibnamefont{and}
  \bibinfo{author}{\bibfnamefont{A.~J.} \bibnamefont{Williamson}},
  \bibinfo{journal}{Physical Review B - Condensed Matter and Materials Physics}
  \textbf{\bibinfo{volume}{71}}, \bibinfo{pages}{1} (\bibinfo{year}{2005}).

\bibitem[{\citenamefont{Alf{\`{e}} and
  Gillan}(2004{\natexlab{b}})}]{Alfe2004-lo}
\bibinfo{author}{\bibfnamefont{D.}~\bibnamefont{Alf{\`{e}}}} \bibnamefont{and}
  \bibinfo{author}{\bibfnamefont{M.~J.} \bibnamefont{Gillan}},
  \bibinfo{journal}{Journal of Physics Condensed Matter}
  \textbf{\bibinfo{volume}{16}}, \bibinfo{pages}{L305}
  (\bibinfo{year}{2004}{\natexlab{b}}).

\bibitem[{\citenamefont{Esler et~al.}(2012)\citenamefont{Esler, Kim, Ceperley,
  and Shulenburger}}]{Esler2012}
\bibinfo{author}{\bibfnamefont{K.}~\bibnamefont{Esler}},
  \bibinfo{author}{\bibfnamefont{J.}~\bibnamefont{Kim}},
  \bibinfo{author}{\bibfnamefont{D.}~\bibnamefont{Ceperley}}, \bibnamefont{and}
  \bibinfo{author}{\bibfnamefont{L.}~\bibnamefont{Shulenburger}},
  \bibinfo{journal}{Computing in Science \& Engineering}
  \textbf{\bibinfo{volume}{14}}, \bibinfo{pages}{40} (\bibinfo{year}{2012}).

\bibitem[{\citenamefont{Krogel and Reboredo}(2018)}]{Krogel2018}
\bibinfo{author}{\bibfnamefont{J.~T.} \bibnamefont{Krogel}} \bibnamefont{and}
  \bibinfo{author}{\bibfnamefont{F.~A.} \bibnamefont{Reboredo}},
  \bibinfo{journal}{Journal of Chemical Physics}
  \textbf{\bibinfo{volume}{148}}, \bibinfo{pages}{044110}
  (\bibinfo{year}{2018}).

\bibitem[{\citenamefont{Dzubak et~al.}(2017{\natexlab{b}})\citenamefont{Dzubak,
  Krogel, and Reboredo}}]{Dzubak2017a}
\bibinfo{author}{\bibfnamefont{A.~L.} \bibnamefont{Dzubak}},
  \bibinfo{author}{\bibfnamefont{J.~T.} \bibnamefont{Krogel}},
  \bibnamefont{and} \bibinfo{author}{\bibfnamefont{F.~A.}
  \bibnamefont{Reboredo}}, \bibinfo{journal}{The Journal of Chemical Physics}
  \textbf{\bibinfo{volume}{147}}, \bibinfo{pages}{024102}
  (\bibinfo{year}{2017}{\natexlab{b}}).

\bibitem[{\citenamefont{Krogel and Kent}(2017)}]{Krogel2017}
\bibinfo{author}{\bibfnamefont{J.~T.} \bibnamefont{Krogel}} \bibnamefont{and}
  \bibinfo{author}{\bibfnamefont{P.~R.~C.} \bibnamefont{Kent}},
  \bibinfo{journal}{The Journal of Chemical Physics}
  \textbf{\bibinfo{volume}{146}}, \bibinfo{pages}{244101}
  (\bibinfo{year}{2017}).

\bibitem[{\citenamefont{Drummond et~al.}(2016)\citenamefont{Drummond, Trail,
  and Needs}}]{Drummond2016}
\bibinfo{author}{\bibfnamefont{N.~D.} \bibnamefont{Drummond}},
  \bibinfo{author}{\bibfnamefont{J.~R.} \bibnamefont{Trail}}, \bibnamefont{and}
  \bibinfo{author}{\bibfnamefont{R.~J.} \bibnamefont{Needs}},
  \bibinfo{journal}{Physical Review B} \textbf{\bibinfo{volume}{94}},
  \bibinfo{pages}{1} (\bibinfo{year}{2016}).

\bibitem[{\citenamefont{Krogel et~al.}(2016)\citenamefont{Krogel, Santana, and
  Reboredo}}]{Krogel2016}
\bibinfo{author}{\bibfnamefont{J.~T.} \bibnamefont{Krogel}},
  \bibinfo{author}{\bibfnamefont{J.~A.} \bibnamefont{Santana}},
  \bibnamefont{and} \bibinfo{author}{\bibfnamefont{F.~A.}
  \bibnamefont{Reboredo}}, \bibinfo{journal}{Physical Review B}
  \textbf{\bibinfo{volume}{93}}, \bibinfo{pages}{075143}
  (\bibinfo{year}{2016}).

\bibitem[{\citenamefont{Trail and Needs}(2017)}]{Trail2017pp}
\bibinfo{author}{\bibfnamefont{J.~R.} \bibnamefont{Trail}} \bibnamefont{and}
  \bibinfo{author}{\bibfnamefont{R.~J.} \bibnamefont{Needs}},
  \bibinfo{journal}{The Journal of Chemical Physics}
  \textbf{\bibinfo{volume}{146}}, \bibinfo{pages}{204107}
  (\bibinfo{year}{2017}).

\bibitem[{\citenamefont{Hohenberg and Kohn}(1964)}]{Hohenberg1964}
\bibinfo{author}{\bibfnamefont{P.}~\bibnamefont{Hohenberg}} \bibnamefont{and}
  \bibinfo{author}{\bibfnamefont{W.}~\bibnamefont{Kohn}},
  \bibinfo{journal}{Phys. Rev.} \textbf{\bibinfo{volume}{136}},
  \bibinfo{pages}{B864} (\bibinfo{year}{1964}).

\bibitem[{\citenamefont{Kohn and Sham}(1965)}]{Kohn1965}
\bibinfo{author}{\bibfnamefont{W.}~\bibnamefont{Kohn}} \bibnamefont{and}
  \bibinfo{author}{\bibfnamefont{L.~J.} \bibnamefont{Sham}},
  \bibinfo{journal}{Physical Review} \textbf{\bibinfo{volume}{140}},
  \bibinfo{pages}{A1133} (\bibinfo{year}{1965}).

\bibitem[{\citenamefont{Kim et~al.}(2018)\citenamefont{Kim, Baczewski, Beaudet,
  Benali, Bennett, Berrill, Blunt, Borda, Casula, Ceperley
  et~al.}}]{qmcpack2018}
\bibinfo{author}{\bibfnamefont{J.}~\bibnamefont{Kim}},
  \bibinfo{author}{\bibfnamefont{A.~T.} \bibnamefont{Baczewski}},
  \bibinfo{author}{\bibfnamefont{T.~D.} \bibnamefont{Beaudet}},
  \bibinfo{author}{\bibfnamefont{A.}~\bibnamefont{Benali}},
  \bibinfo{author}{\bibfnamefont{M.~C.} \bibnamefont{Bennett}},
  \bibinfo{author}{\bibfnamefont{M.~A.} \bibnamefont{Berrill}},
  \bibinfo{author}{\bibfnamefont{N.~S.} \bibnamefont{Blunt}},
  \bibinfo{author}{\bibfnamefont{E.~J.~L.} \bibnamefont{Borda}},
  \bibinfo{author}{\bibfnamefont{M.}~\bibnamefont{Casula}},
  \bibinfo{author}{\bibfnamefont{D.~M.} \bibnamefont{Ceperley}},
  \bibnamefont{et~al.}, \bibinfo{journal}{Journal of Physics: Condensed Matter}
  \textbf{\bibinfo{volume}{30}}, \bibinfo{pages}{195901}
  (\bibinfo{year}{2018}).

\bibitem[{\citenamefont{Martin}(2004)}]{MartinBook2004}
\bibinfo{author}{\bibfnamefont{R.~M.} \bibnamefont{Martin}},
  \emph{\bibinfo{title}{Electronic {Structure}: {Basic} {Theory} and
  {Practical} {Methods}}} (\bibinfo{publisher}{Cambridge University Press},
  \bibinfo{year}{2004}), ISBN \bibinfo{isbn}{0-521-78285-6}.

\bibitem[{\citenamefont{Root et~al.}(2015)\citenamefont{Root, Shulenburger,
  Lemke, Dolan, Mattsson, and Desjarlais}}]{Root2015}
\bibinfo{author}{\bibfnamefont{S.}~\bibnamefont{Root}},
  \bibinfo{author}{\bibfnamefont{L.}~\bibnamefont{Shulenburger}},
  \bibinfo{author}{\bibfnamefont{R.~W.} \bibnamefont{Lemke}},
  \bibinfo{author}{\bibfnamefont{D.~H.} \bibnamefont{Dolan}},
  \bibinfo{author}{\bibfnamefont{T.~R.} \bibnamefont{Mattsson}},
  \bibnamefont{and} \bibinfo{author}{\bibfnamefont{M.~P.}
  \bibnamefont{Desjarlais}}, \bibinfo{journal}{Physical Review Letters}
  \textbf{\bibinfo{volume}{115}}, \bibinfo{pages}{1} (\bibinfo{year}{2015}).

\bibitem[{\citenamefont{Giannozzi et~al.}(2009)\citenamefont{Giannozzi, Baroni,
  Bonini, Calandra, Car, Cavazzoni, Ceresoli, Chiarotti, Cococcioni, Dabo
  et~al.}}]{qe:main}
\bibinfo{author}{\bibfnamefont{P.}~\bibnamefont{Giannozzi}},
  \bibinfo{author}{\bibfnamefont{S.}~\bibnamefont{Baroni}},
  \bibinfo{author}{\bibfnamefont{N.}~\bibnamefont{Bonini}},
  \bibinfo{author}{\bibfnamefont{M.}~\bibnamefont{Calandra}},
  \bibinfo{author}{\bibfnamefont{R.}~\bibnamefont{Car}},
  \bibinfo{author}{\bibfnamefont{C.}~\bibnamefont{Cavazzoni}},
  \bibinfo{author}{\bibfnamefont{D.}~\bibnamefont{Ceresoli}},
  \bibinfo{author}{\bibfnamefont{G.~L.} \bibnamefont{Chiarotti}},
  \bibinfo{author}{\bibfnamefont{M.}~\bibnamefont{Cococcioni}},
  \bibinfo{author}{\bibfnamefont{I.}~\bibnamefont{Dabo}}, \bibnamefont{et~al.},
  \bibinfo{journal}{Journal of physics. Condensed matter}
  \textbf{\bibinfo{volume}{21}}, \bibinfo{pages}{395502}
  (\bibinfo{year}{2009}).

\bibitem[{\citenamefont{Giannozzi et~al.}(2017)\citenamefont{Giannozzi,
  Andreussi, Brumme, Bunau, {Buongiorno Nardelli}, Calandra, Car, Cavazzoni,
  Ceresoli, Cococcioni et~al.}}]{Giannozzi2017}
\bibinfo{author}{\bibfnamefont{P.}~\bibnamefont{Giannozzi}},
  \bibinfo{author}{\bibfnamefont{O.}~\bibnamefont{Andreussi}},
  \bibinfo{author}{\bibfnamefont{T.}~\bibnamefont{Brumme}},
  \bibinfo{author}{\bibfnamefont{O.}~\bibnamefont{Bunau}},
  \bibinfo{author}{\bibfnamefont{M.}~\bibnamefont{{Buongiorno Nardelli}}},
  \bibinfo{author}{\bibfnamefont{M.}~\bibnamefont{Calandra}},
  \bibinfo{author}{\bibfnamefont{R.}~\bibnamefont{Car}},
  \bibinfo{author}{\bibfnamefont{C.}~\bibnamefont{Cavazzoni}},
  \bibinfo{author}{\bibfnamefont{D.}~\bibnamefont{Ceresoli}},
  \bibinfo{author}{\bibfnamefont{M.}~\bibnamefont{Cococcioni}},
  \bibnamefont{et~al.}, \bibinfo{journal}{Journal of Physics: Condensed Matter}
  \textbf{\bibinfo{volume}{29}}, \bibinfo{pages}{465901}
  (\bibinfo{year}{2017}).

\bibitem[{\citenamefont{Umrigar et~al.}(2007)\citenamefont{Umrigar, Toulouse,
  Filippi, Sorella, and Hennig}}]{Umrigar:linear}
\bibinfo{author}{\bibfnamefont{C.~J.} \bibnamefont{Umrigar}},
  \bibinfo{author}{\bibfnamefont{J.}~\bibnamefont{Toulouse}},
  \bibinfo{author}{\bibfnamefont{C.}~\bibnamefont{Filippi}},
  \bibinfo{author}{\bibfnamefont{S.}~\bibnamefont{Sorella}}, \bibnamefont{and}
  \bibinfo{author}{\bibfnamefont{R.~G.} \bibnamefont{Hennig}},
  \bibinfo{journal}{Physical Review Letters} \textbf{\bibinfo{volume}{98}},
  \bibinfo{pages}{110201} (\bibinfo{year}{2007}).

\bibitem[{\citenamefont{Mathuriya
  et~al.}(2017{\natexlab{a}})\citenamefont{Mathuriya, Luo, Benali,
  Shulenburger, and Kim}}]{ipdps}
\bibinfo{author}{\bibfnamefont{A.}~\bibnamefont{Mathuriya}},
  \bibinfo{author}{\bibfnamefont{Y.}~\bibnamefont{Luo}},
  \bibinfo{author}{\bibfnamefont{A.}~\bibnamefont{Benali}},
  \bibinfo{author}{\bibfnamefont{L.}~\bibnamefont{Shulenburger}},
  \bibnamefont{and} \bibinfo{author}{\bibfnamefont{J.}~\bibnamefont{Kim}}, in
  \emph{\bibinfo{booktitle}{Parallel and Distributed Processing Symposium
  (IPDPS), 2017 IEEE International}} (\bibinfo{organization}{IEEE},
  \bibinfo{year}{2017}{\natexlab{a}}), pp. \bibinfo{pages}{213--223}.

\bibitem[{\citenamefont{Mathuriya
  et~al.}(2017{\natexlab{b}})\citenamefont{Mathuriya, Luo, Clay, Benali,
  Shulenburger, and Kim}}]{supercomputing}
\bibinfo{author}{\bibfnamefont{A.}~\bibnamefont{Mathuriya}},
  \bibinfo{author}{\bibfnamefont{Y.}~\bibnamefont{Luo}},
  \bibinfo{author}{\bibfnamefont{R.~C.} \bibnamefont{Clay},
  \bibfnamefont{III}},
  \bibinfo{author}{\bibfnamefont{A.}~\bibnamefont{Benali}},
  \bibinfo{author}{\bibfnamefont{L.}~\bibnamefont{Shulenburger}},
  \bibnamefont{and} \bibinfo{author}{\bibfnamefont{J.}~\bibnamefont{Kim}}, in
  \emph{\bibinfo{booktitle}{Proceedings of the International Conference for
  High Performance Computing, Networking, Storage and Analysis}}
  (\bibinfo{publisher}{ACM}, \bibinfo{address}{New York, NY, USA},
  \bibinfo{year}{2017}{\natexlab{b}}), SC '17, pp.
  \bibinfo{pages}{38:1--38:12}, ISBN \bibinfo{isbn}{978-1-4503-5114-0}.

\end{thebibliography}
\end{document}